\documentclass[journal]{IEEEtran}
\usepackage{graphicx}
\usepackage[noadjust]{cite}
\usepackage{enumitem}

\usepackage[cmex10]{amsmath}

\usepackage{amssymb}
\usepackage{caption}
\usepackage{wasysym}
\usepackage{soul}
\usepackage{mathtools,cuted}
\usepackage{algorithmic}
\usepackage{algorithm}

\usepackage{algorithmic}
\usepackage{subfigure}
\usepackage[section]{placeins}

\usepackage{array}
\usepackage{multirow}

\usepackage{stfloats}

\usepackage{booktabs}

\usepackage{float}
\usepackage{hyperref}

\usepackage{url}

\begin{document}

\title{FPGA based design for online computation of Multivariate EMD (MEMD)}



\author{\IEEEauthorblockN{Sikender Gul,
		Muhammad Faisal Siddiqui, and
		Naveed Ur Rehman }
	
	\thanks{ The authors are affiliated with the department of Electrical and Computer Engineering, COMSATS University Islamabad, Pakistan. Corresponding author: Naveed Ur Rehman (email:naveed.rehman@comsats.edu.pk)}}

\markboth{}%
{Shell \MakeLowercase{\textit{et al.}}: Bare Demo of IEEEtran.cls for IEEE Transactions on Magnetics Journals}
%




\IEEEtitleabstractindextext{%
\begin{abstract}
Multivariate or multichannel data have become ubiquitous in many modern scientific and engineering applications, e.g., biomedical engineering, owing to recent advances in sensor and computing technology. Processing these data sets is challenging owing to: i) their large size and multidimensional nature, thus requiring specialized algorithms and efficient hardware designs for on-line and real-time processing; ii) the nonstationary nature of data arising in many real life applications demanding new extensions of standard multiscale non-stationary signal processing tools. In this paper, we address the former issue by proposing a fully FPGA based hardware architecture of a popular multi-scale and multivariate signal processing algorithm, termed as multivariate empirical mode decomposition (MEMD). MEMD is a data-driven method that extends the functionality of standard empirical mode decomposition (EMD) algorithm to multichannel or multivariate data sets. Since its inception in 2010, the algorithm has found wide spread applications spanning different engineering related fields. Yet, no parallel FPGA based hardware design of the algorithm is available for its on-line and real-time processing. Our proposed architecture for MEMD uses fixed point operations and employs cubic spline interpolation within the sifting process. Finally, examples of decomposition of multivariate synthetic and real world biological signals are provided.
\end{abstract}

\begin{IEEEkeywords}
Multivariate signals, multivariate empirical mode decomposition, FPGA, time frequency methods
\end{IEEEkeywords}}

\maketitle
\IEEEdisplaynontitleabstractindextext
\IEEEpeerreviewmaketitle

\section{Introduction}
Empirical mode decomposition (EMD) \cite{1} is a data-driven technique that is widely used for the decomposition and time-frequency (T-F) analysis of nonlinear and non-stationary signals. Unlike traditional multi-scale data analysis techniques, such as Fourier and Wavelet
transform, that use fixed a priori basis function for decomposition, EMD effectively employs local data adaptive basis functions for decomposition of data. To achieve that, EMD adopts an iterative sifting process to decomposes signal into its multiple inherent scales with distinct features, known as intrinsic mode functions (IMFs). The IMFs are designed such that they are zero-mean and observe well-behaved Hilbert spectrum, termed as Hilbert-Huang spectrum.  

The EMD and its associated HHS are extensively used in many applications related to nonstationary signal processing, such as biomedical signal processing \cite{x2}, \cite{5}, fault detection in mechanical gearboxes \cite{x4}, detection of low frequency oscillation \cite{x5}, data fusion \cite{x6}, and data denoising \cite{x7}. In its original formulation, EMD can only cater for univariate or single-channel data sets. In real life applications, however, we often encounter nonstationary signals containing multiple data channels, e.g., as in \cite{X8}, \cite{x9}, \cite{X10}, and \cite{x11}. EMD being a natural candidate for the processing of nonstationary data was also considered for the processing of multichannel data sets. However, a typical approach was to apply EMD on each channel of a multivariate data in isolation. That is a naive approach though since it disregards interchannel correlations within multivariate data, resulting in well known frequency-alignment problems within EMD \cite{x16b}. 

To resolve that problem, a fully multivariate extension of EMD was proposed in \cite{x12} that operates on a multivariate signal directly in multidimensional space where it resides. The algorithm is termed as multivariate empirical mode decomposition (MEMD). It uses a sifting process, similar to EMD, for decomposition of multivariate signals into multidimensional IMFs. Since its inception, the method has found cross disciplinary applications in wide ranging fields including biomedical signal classification and related applications \cite{x13}, \cite{x15}, fault diagnosis in machines \cite{X10}, data fusion {x25b}, process control \cite{x14} and data denoising \cite{den}. 

Despite their success, EMD and its related algorithms are yet to make a big impact in applications involving real-time and on-line processing of data. There are two main reasons for that: i) scarcity of algorithms for on-line computation of EMD and its extensions; ii) lack of efficient hardware architectures for EMD computation. Those problems could be mainly attributed to the recursive nature of the EMD related methods that makes hardware resource utilization, such as in FPGA based designs, a difficult task. Still, efficient hardware based designs for on-line EMD and its multivariate extensions are prerequisites for utilization of these algorithms in a number of applications spanning different fields.   

In literature, several parallel and pipelined based hardware implementations of EMD \cite{17}, \cite{18}, \cite{19}, \cite{20}, \cite{x24}, and \cite{21} are available. For general multivariate extension of EMD (MEMD), however, only software implementations; i.e., based on MATLAB and C programming language; or GPU based designs are available that limits its usage exclusively for off-line applications. Specifically, the GPU based implementation for MEMD,  using the compute unified device architecture (CUDA) architecture, was presented in \cite{x20} that exploits inherent parallelism within the MEMD algorithm. The parallel design improved the speed by up to seven times when compared to C programming based serial implementation of MEMD. However, the design only works on batch data thereby limited its applications to off-line applications. Moreover, {the execution time to decompose 1000 samples is 750$m$s which is very high for real-time application.} 

For multivariate signals with two channels only (bi-variate signals), the FPGA-based parallel design for computation of {bi-variate EMD \cite{x28}} was proposed in \cite{21}. The design is implemented on Xilinx Kintex 7 FPGA and can operate up to maximum operating frequency of 24 MHz. However, it uses linear interpolation instead of more established cubic spline interpolation (CSI) to implement the sifting process that compromises its accuracy. 

There are currently no FPGA based fully hardware architectures for computing the multivariate EMD (MEMD) algorithm. Given the massive appeal of the MEMD method in real life multichannel data sets, we propose the first-ever fully FPGA based fixed-point architecture for multivariate EMD algorithm. The proposed design uses accurate cubic spline interpolation (CSI) scheme for the sifting process within MEMD thus fully ensuring the accuracy of the implementation.{ Moreover, the parallel structure of the proposed design enables processing of multiple channels of input data simultaneously, resulting in higher throughput of the system.} We present a detailed timing analysis, hardware resource utilization of proposed architecture in addition to a thorough comparison with the state-of-the-art. The design is verified and evaluated on a number of synthetic and real life multivariate signals. 

The rest of paper is organized as follows: Section II reviews the MEMD algorithm. The implementation of proposed design is presented in Section III. The experimental result of synthetic and real world data sets with discussion is presented in section IV. Finally, Section V presents the  conclusions and future avenues for research in this area.

\section{REVIEW OF MEMD ALGORITHM}\label{sec:MEMD_rev}
MEMD is multivariate extension of the EMD algorithm that has gained remarkable success in applications involving multichannel data processing e.g. image fusion \cite{x25}, biomedical engineering \cite{x26} and condition monitoring \cite{x26b}.
MEMD decomposes multivarite signal in to its intrinsic multiply oscillatory modes, known as multivariate intrinsic mode function (IMFs). MEMD uses the sifting process to decomposes input multivariate data $\bar{X}(t)$ into multivariate finite no of IMFs $\bar{C_j}(t)$, written as:
\begin{equation}
\bar{X}(t)=\sum_{j=1}^{M}\bar{C_j}(t)+\bar{r}(t).
\label{eq:MEMD}
\end{equation}

where $M$ is the number of decomposed IMFs from the multivariate input data sets and $\bar{r}(t)$ is residue signal. The IMFs $\bar{C_j}(t)$ is defined by function that satisfy the following two conditions:

\begin{enumerate}[label=\Roman*.]
	\item The number of zero-crossings and extrema points must either equal or differ at most by one.
	\item The local mean envelopes must be equal to zero which are defined by local maxima and local minima.
\end{enumerate}

\begin{figure}[t]
	\centering
	\captionsetup{justification=centering}
	\includegraphics[height=2.0in,width=3.5in,trim=2mm 2mm 4mm 4mm, clip=true]{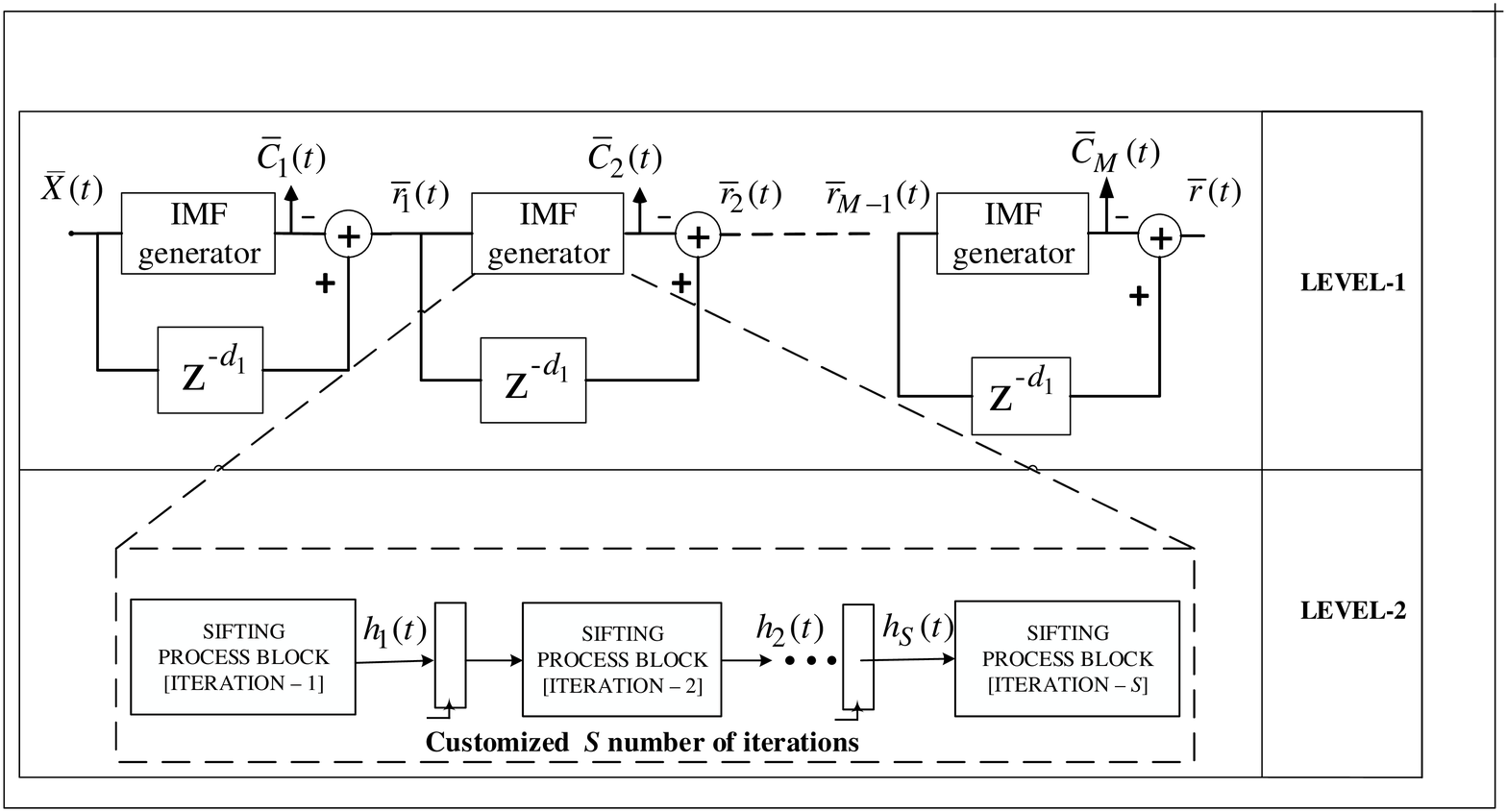}
	\caption{Overall block diagram of the proposed MEMD design.}
	\label{fig1}       
\end{figure}

\begin{algorithm}[h!]
	\caption{\bf The Sifting Process in MEMD \cite{x12}}
	\label{alg:EMD}
	\begin{algorithmic}[1]
		\STATE Generate the point set using Hamersley sequence that used for sampling a $n(n-1)$ sphere.
		\STATE Calculate a projection of the  input signal $\bar{X}(t)$ along the direction vector $K$, giving as the set of direction vector.
		\STATE Locate  the extrema (maxima and minima) point  from the set of projected signal and its corresponding time instants.
		
		\STATE Interpolate extrema point corresponding time instants to acquire multivariate envelope curves.
		
		\STATE 	The local mean $m(t)$ of multiple envelopes for $K$ direction vector is given by:
		\begin{equation}
		m\left(t\right)=\frac{1}{K}\sum_{n=1}^{K}{e^{\theta_n}(t)}
		\end{equation}

		\STATE The oscillatory mode $\bar{h}(t)$ is obtain by subtracting local mean $m(t)$ from the input signal
		\begin{equation}
		\bar{h}(t)=\bar{X}(t)-m(t);\nonumber
		\end{equation}
		\STATE If $\bar{h}(t)$ meets the stopping criteria as stated above, then define
		 $\bar{C_j}=\bar{h}(t)$ as an IMF, otherwise set $\bar{X}(t)= \bar{h}(t)$ and repeat the procedures from step 1.
	\end{algorithmic}
\end{algorithm}

 In standard  EMD, the local mean is calculated by averaging the upper and lower envelope. However, the local mean of multi-dimensional signals cannot be generated like as in standard EMD. In MEMD, multiple n-dimensional envelopes are calculated by projecting the multi-dimensional signal along different directions in n-dimensional space and the corresponding local means are then calculated by averaging the envelopes from those projections. Hammersley sequences are used to acquire quasi-uniform points on high dimensional spheres \cite{x22}, \cite{x23} for uniform set of direction vector. The details of MEMD Algorithm is presented in Algorithm 1. The local mean of  MEMD is estimated by computing the multiple real-valued projections and taking the local mean of corresponding envelopes.


\section{FPGA BASED ARCHITECTURE FOR multivariate EMD}\label{sec:arc}

We exploit the inherent parallelism of the MEMD algorithm to efficiently implement it on FPGA using parallel and pipeline structure. {The proposed design employs both fixed-point format and CSI within sifting process of MEMD}. FPGA-based design for the on-line computation of multivariate extension of EMD (MEMD) is presented in this section. The overall block diagram of the proposed architecture is presented first and then subsequent individual modules are discussed in detail.

\subsection{Architectural design of multivariate extension of EMD  } \label{subsec:arc}

Fig. \ref{fig1} Shows the operational flow of the proposed hardware design for the computation of MEMD. The block diagram is divided into two levels. The upper portion of Fig. \ref{fig1} shows level 1 where different IMFs are calculated using pipeline architecture. At level 2, multiple iteration blocks are cascaded in the pipeline as shown in the lower portion of Fig. \ref{fig1}, where the sifting process is performed. 

The input signal $\bar{X}(t)$ denotes a 16-bit multichannel data which is fed to the multivariate IMF generator block to generate the first IMF $\bar{C_{1}}(t)$. The IMF generator block also  produces the multi-dimensional residue $\bar{r_{1}}(t)$, by subtracting $\bar{C_{1}}(t)$ from the input signal $\bar{X}(t)$. The residue $\bar{r_{1}}(t)$ is then fed to the second stage of IMF generator as an input to produce the second IMF $\bar{C_{2}}(t)$. In our design, the $M$ number of IMF generator blocks are cascaded to produce the $M$ number of IMFs and a residue. This is flexible as multiple IMF generator blocks can be cascaded depending on the number of required IMFs. 

The lower portion of Fig. \ref{fig1} shows the IMF generator block where sifting process is performed by using a pipelined structure. Note that $S$ number of sifting blocks are used in our design for obtaining a single IMF, in accordance with the $S$-number stopping criterion~\cite{24} of EMD. The first sifting operation produces the output $h_1(t)$ that becomes the input to the second iteration block for the computation of $h_2(t)$. Similarly, the output of the second iteration, $h_2(t)$, becomes  input to the third iteration block and so on. The output $h_S(t)$ of the last sifting block becomes the first IMF from the sifting process i.e., $\bar{C_{1}}(t)=h_S(t)$.

\begin{figure*}[t]
	\centering
	\captionsetup{justification=centering}
	\includegraphics[height=2.7in,width=7in,trim=0mm 0mm 0mm 0mm, clip=true]{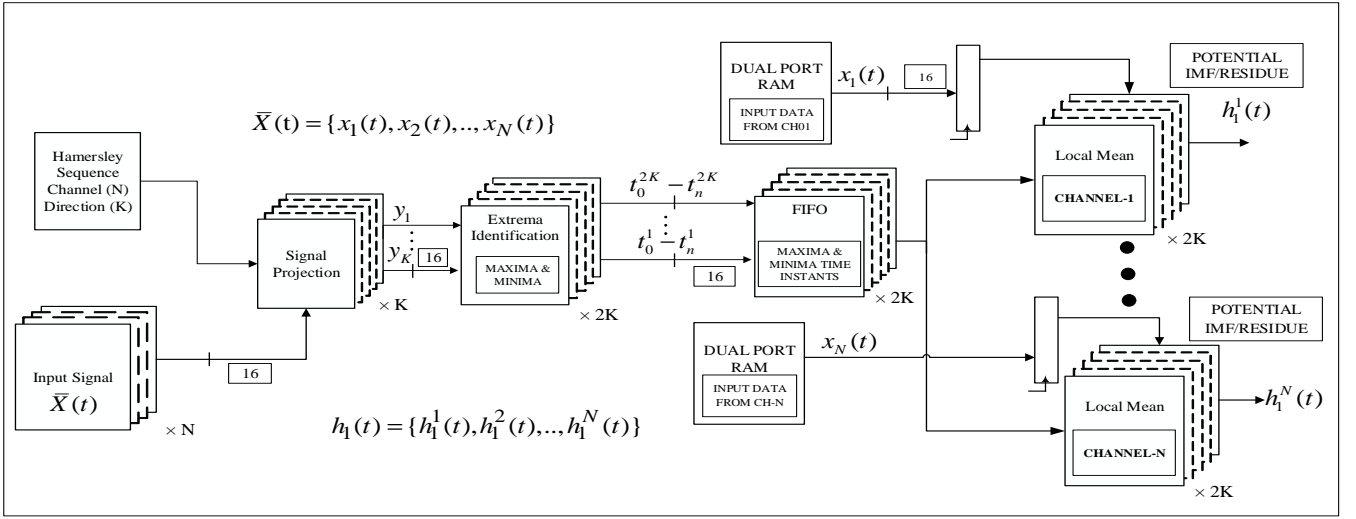}
	\caption{Representation of the sifting process within the proposed MEMD architecture.}
	\label{fig1a}       
\end{figure*}

\subsection{Sifting process} \label{subsec:sifting}

The sifting process is the core component for the computation of MEMD. Sifting process involves multiple iteration blocks to compute each IMF. The proposed architecture for a single iteration block to calculate the sifting process within MEMD is shown in Fig.\ref{fig1a}. The iteration block is composed of a signal projection module, Hammersley sequence generator, extrema identification module, local mean module (for each channel) and different memory units for data storage. 

The sifting process starts with the calculation of $K$ number of direction vectors by using pre-generated Hammersley sequence. Hammersley sequence is pre-generated and stored on the block memory of FPGA. The input signal $\bar{X}(t)$ is concurrently stored on dual-port block RAM. The signal projection module is next utilized to project $\bar{X}(t)$ along $K$ direction vectors to compute the projected signals $\{y_1(t),y_2(t),...,y_k(t)\}$.

\subsubsection[Projection]{Signal Projection}
\begin{figure*}[t]
	\centering
	\captionsetup{justification=centering}
	\includegraphics[scale=0.4,trim=2mm 25mm 2mm 2mm, clip=true]{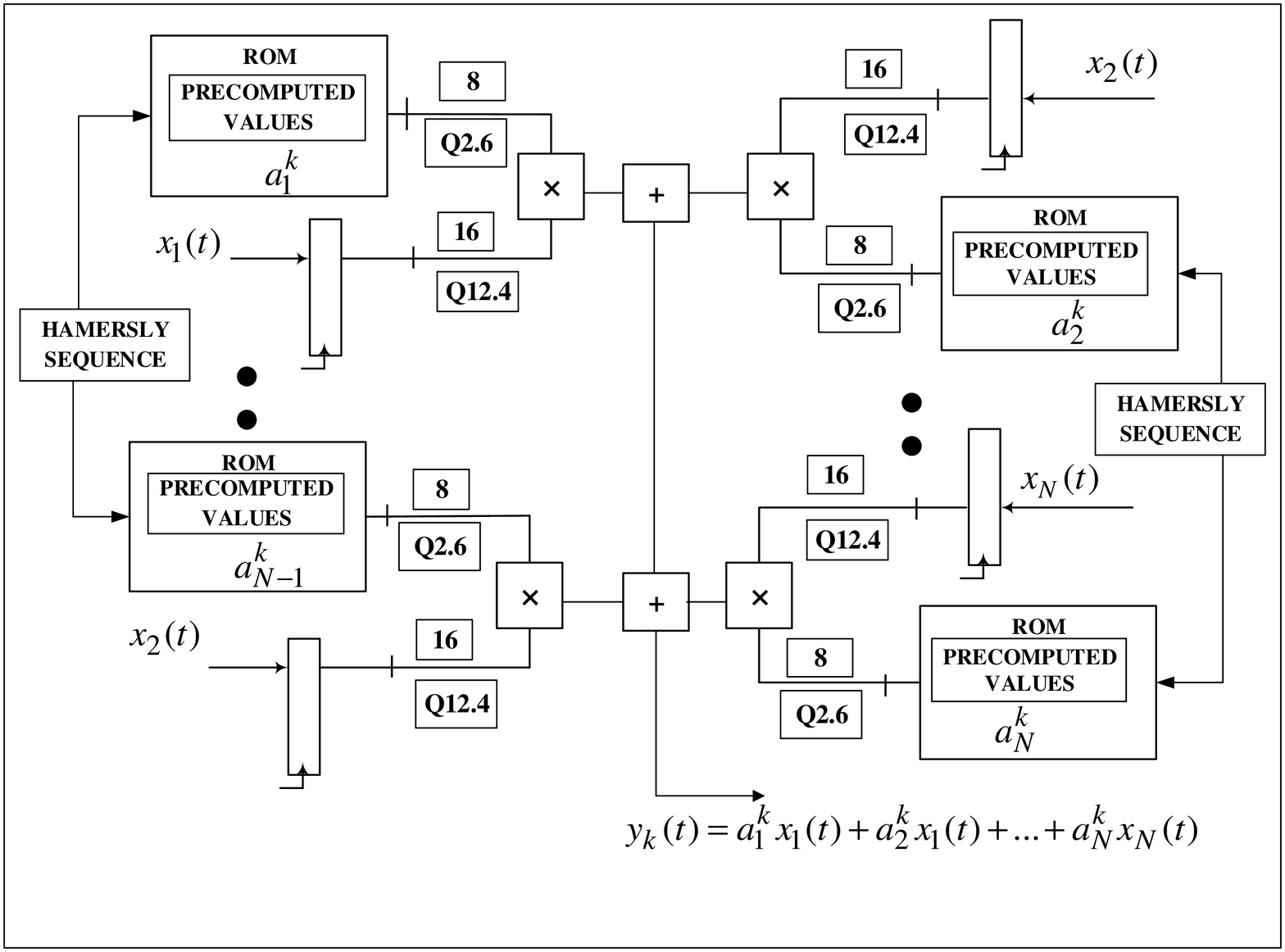}
	\caption{Block diagram of the signal projection module.}
	\label{fig1_projection}       
\end{figure*}

The signal projection module is shown in Fig. \ref{fig1_projection}. 
It uses Hammersly sequence which is pre-calculated and stored on a ROM. The input signal $x_i\}_{i=1}^{N}$ is projected along multiple direction vectors to generate 
$K$ number of projected signals. The $k-$th such projection is given by: 
\begin{equation}
y_k(t)=a_1^k x_1(t)+a_{2}^k x_2(t)+ ....+a_{N}^k x_N
\label{eq2_x}
\end{equation}
where $a_{ik}$ are the coefficients of the Hammerseley sequence that are used to obtain the signal projection along the $k-$th direction. 

In Fig. \ref{fig1_projection}, the architecture used for the computation of \eqref{eq2_x} is shown. As the coefficients of Hammerseley sequence $a_{i}^{k}$ are fixed values, they are stored in the ROM. We employ canonical signed digit (CSD) based multiplication within the signal projection module as the coefficients of the Hammerseley sequence are constants. The multiplication with the constant is replaced by shift operations and additions within CSD multiplication to increase the throughput of the system.

\subsubsection{Extrema identification}
\begin{figure}[t]
	\centering
	\captionsetup{justification=centering}
	\includegraphics[scale=0.35,trim=2mm 2mm 2mm 5mm, clip=true]{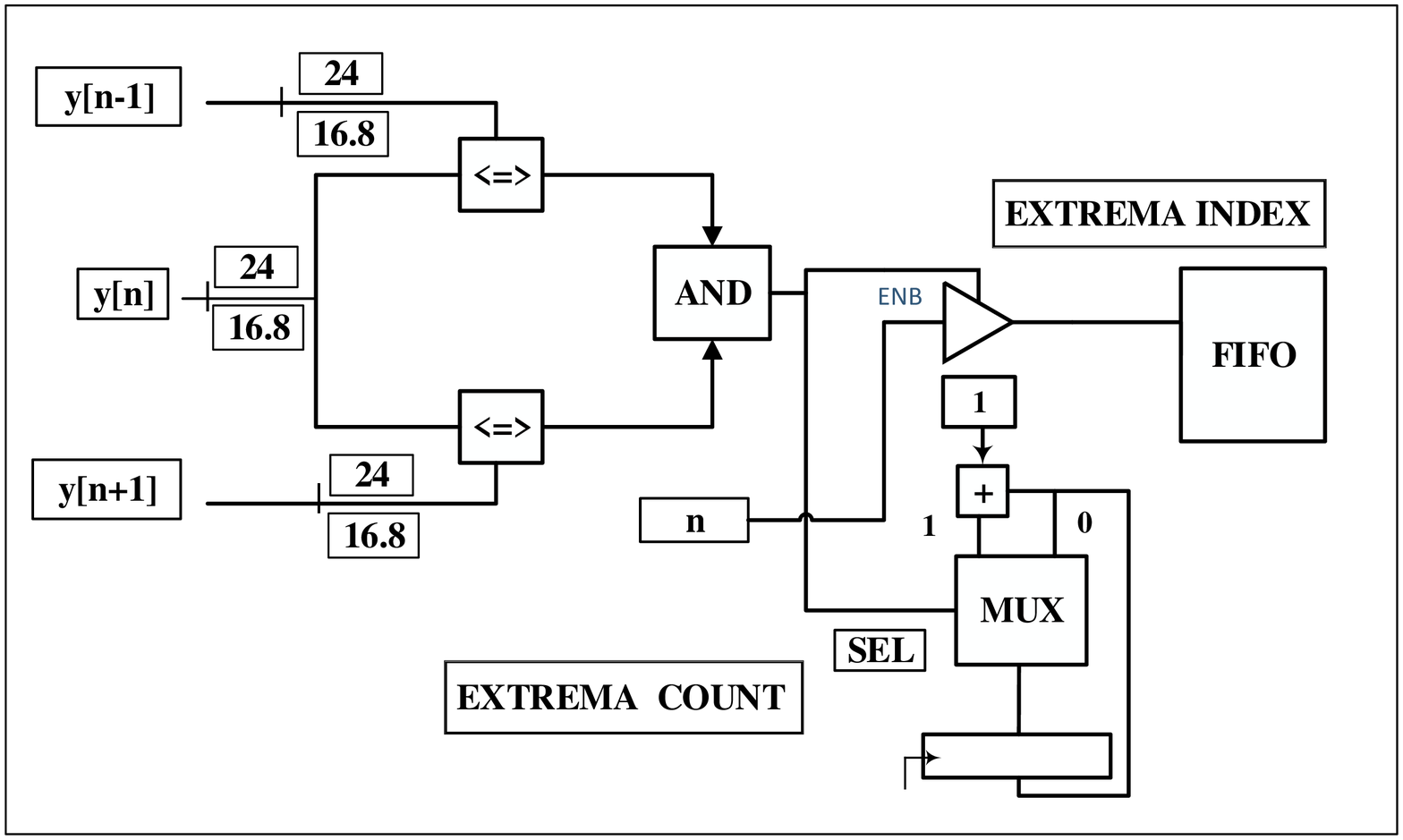}
	\caption{Block diagram of the extrema identification module.}
	\label{fig1_EI}       
\end{figure}

 The sifting process of MEMD involves envelope generation for each projected signal $y_k(t)$. To achieve that, extrema values of the projected signals are detected along with their corresponding time instants. 
 Both sets of values are stored in FIFOs. The extrema points are detected using the extrema identification module, as shown in Fig. \ref{fig1_EI}. The extrema identification module takes three input data values $y_{n-1},y_n$ and $y_{n+1}$ to detect an extremum. These input values are sequentially loaded to extrema identification module at every clock cycle. 
 
Two comparators are used to compare $y_{n}$ with $y_{n-1}$ and $y_{n+1}$ separately. If $y_n$ is greater than or equal to $y_{n-1}$, the first comparator generates a high signal. Similarly, the second comparator generates a high signal whenever $y_n$ is greater than or equal to $y_{n+1}$. The outputs from both comparators become input to an AND gate that generates a select (SEL) signal. The SEL signal performs two operations: i) it enables the extrema counter to increment the value by 1; ii) it enables the tri-state buffer to store the value of time indices ${n}$ of extrema in the FIFO. 
 
Note that the same module is used to detect both minimum and maximum values of the projected signal by replacing the condition of the comparator block i.e., $\leq$ gives minimum values whereas $\geq$ gives the maximum values. As the projected signal have both extrema as well as minima, therefore, $2K$ number of extrema identification modules are used in the proposed architecture. The time indices of extrema are further used in local mean module for calculation of local mean, which is explained next. 
 

  
\subsection{Local Mean Generation} \label{subsec:local mean}

The local mean of a multivariate signal is generated based on multiple signal envelopes in $K$ number of directions and for $N$ number of input channels. In the proposed design, the local mean is calculated in parallel as shown in right hand side of Fig\ref{fig1a}. The local mean block for each channel is connected in parallel to calculate $h_1(t)$, a potential candidate for IMF. Therefore, the $N$ number of local mean blocks are used to calculate $h_1(t)$ as follows:
\begin{equation}
h_1(t)=\{h_1^1(t),h_1^1(t),...,h_1^N(t)\}
\label{eq2_y}
\end{equation}
where $h_1^1(t)$ is the local mean of channel 1 and $h_1^N(t)$ is the local mean corresponding to the channel $N$.

\begin{figure}[t]
	\centering
	\captionsetup{justification=centering}
	\includegraphics[scale=0.27,trim=5mm 4mm 5mm 5mm, clip=true]{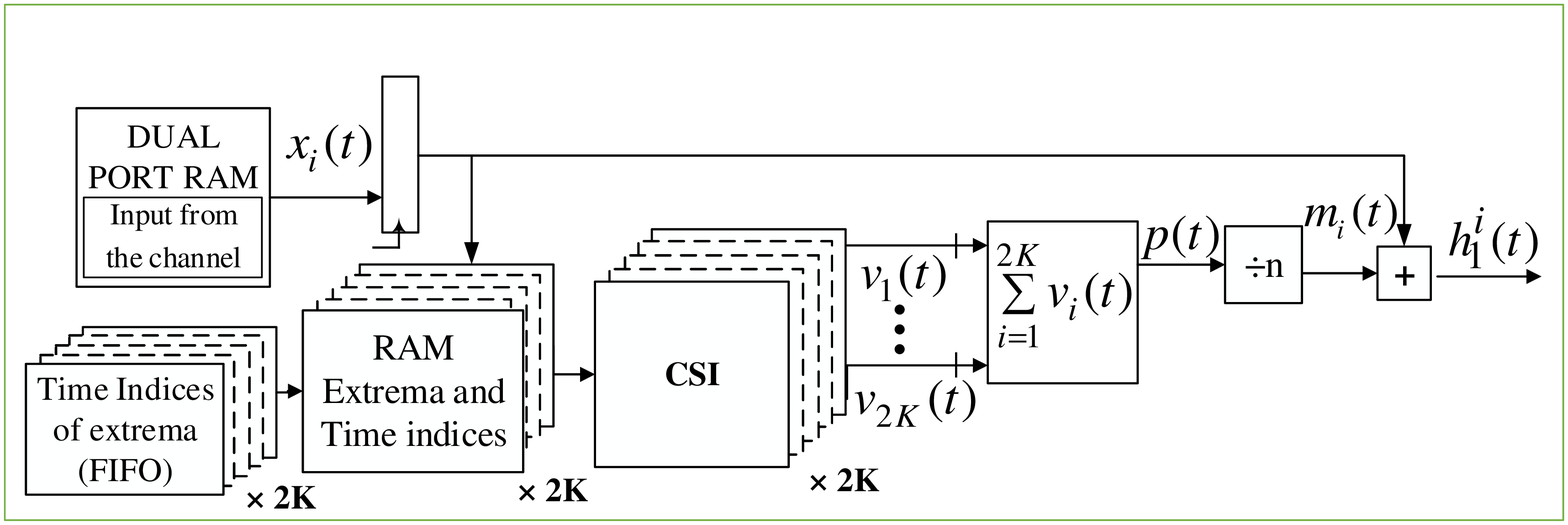}
	\caption{Local mean estimation block.}
	\label{fig1_b}       
\end{figure}

The proposed design to compute local mean $h_1^i(t)$ corresponding to the $i-th$ channel is presented in Fig. \ref{fig1_b}. The $N$ number of such blocks are required to compute the local mean $h_1(t)$. To compute $h_1^i(t)$, multiple signal envelopes in $K$ number of directions are required. In order to generate those envelopes, firstly, time instants of each projected signals' extrema are taken from the FIFO and corresponding extrema values are extracted from the input data $x_i(t)$ of $i-th$ channel which is stored in RAM. The extrema value and their associated position for $K$ number of projected signals are stored in $2K$ number of dual-port RAMs. These extrema values and their associated indices are further used to calculate multiple $2K$ number of signal envelopes, denoted by $V_1(t),V_2(t)..V_{2K}(t)$, by using cubic spline interpolation (CSI) module that will be discussed in the next section. These multiple signal envelopes are then used to calculate mean $m_i(t)$ of $i-th$ channel through the following relation:
\begin{equation}
\begin{aligned}
&m_i(t)=\frac{1}{2K}\sum_{i=1}^{2K}{V_i(t)}.
\end{aligned}
\label{eq3_mean}
\end{equation}

The local mean for a single channel is finally calculated by subtracting mean $m_i(t)$ from the input signal i.e. $h_1^i(t) = x_i(t)- m_i(t)$. Similarly, $N$ numbers of local mean are calculated using \eqref{eq2_y} as the local mean module is cascaded in parallel for each channel to complete the sifting process within MEMD.

\subsection{CSI hardware implementation } \label{subsec:csi_prop}

Cubic spline interpolation (CSI) is one of the most computationally extensive task within any EMD based algorithm. That also applies to MEMD. Therefore, for real time on-line implementation of the MEMD algorithm, we have used the CSI design from our previous published work \cite{x24}. It required only three extremas points and their indices for the formation of spline. The cubic spline operation involves third degree powers of the input parameters and involves four coefficients, as shown in the equation below  \cite{x24}: 

\begin{equation} \label{eq8}
\begin{aligned}
q_j(x)=a_j+b_j(x-x_j)+c_j(x-x_j)^2+d_j(x-x_j)^3.
\end{aligned}
\end{equation}

\noindent where the coefficients $a_j,{\ }b_j,{\ }c_j$ and $d_j$ are given by 

\begin{equation} \label{eq_9}
\begin{aligned}
&a_j=f(x_j)=f_i,\\
&c_j=k_j,\\
&b_j=\frac{{1}}{h_j}(a_{j+1}-a_j){-}\frac{h_j}{3}({2}c_j{+}c_{j+1}),\\
&d_j=\frac{c_{j+1}+c_j}{{3}h_j}.
\end{aligned}
\end{equation}

Note that $h_j=x_{j+1}-x_j$, $f_i$ is the value of function $f$ at point $x_j$ and $k_j$ denotes the derivative of the spline function. It is calculated by using free or natural conditions of the cubic spline and then solved by using tridiagonal matrix algorithm (TDMA), also called the Thomas algorithm~\cite{27}. 

\begin{figure}[t]
	\centering
	\captionsetup{justification=centering}
	\includegraphics[scale=0.275,trim=2mm 5mm 5mm 5mm,clip=true]{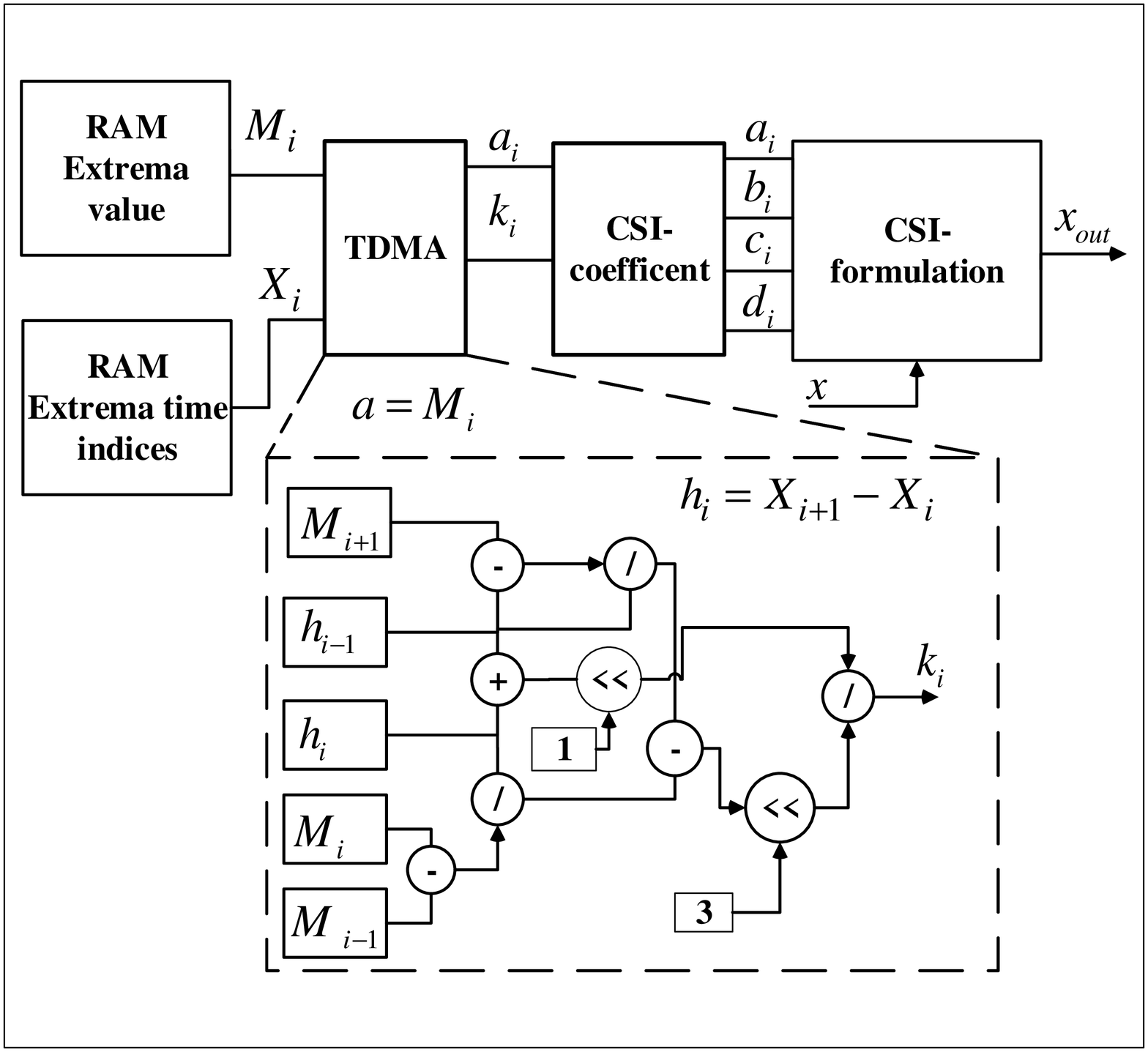}
	\caption{Block diagram of our design for implementing cubic spline interpolation (CSI) process.}
	\label{fig3}       
\end{figure}

The architectural block diagram of the CSI design module is shown in Fig. \ref{fig3}. The CSI block consists of a tridiagonal matrix algorithm sub-block (TDMA) in addition to the CSI-coefficent and CSI-formulation modules. The block takes three extrema values $M_{i-1}, M{i}$ and $M_{i+1}$ along with their indices $X_{i-1}, X_i$ and $X_{i+1}$ from the dual port RAM to calculate the spline according to (6). 

The Tridiagonal matrix algorithm is used to calculate the $a_i$ and $k_i$. The lower portion of the Fig. \ref{fig3} calculates $k_i$ by using \eqref{eq_9}. The division operation in the \eqref{eq_9} is implemented through a look-up table. Since $h_i$ is always a positive number, therefore, a small-sized look-up table can be used to perform the division operation, which decreases the execution time of the proposed design. The CSI-coefficient module is used to calculates $a_j,\mathrm{\ }b_j,\mathrm{\ }c_j,\mathrm{\ }d_j$, which are the coefficient of the spline, via \eqref{eq_9}. The CSI-formation block uses the above coefficient along with $x\mathrm{\ }$ to perform the cubic spline interpolation. {Here, $x$ indicates the time indices of the envelope that need to be interpolated.} Our architecture computes on-lines MEMD, therefore, once the CSI process is completed for a certain signal interval, the next three extrema value are taken from the RAM to generate the signal envelope for those extrema values. The process is repeated until the $K$ signal envelopes are completed.

\section{Results and Analysis}

In this section, we have reported the performance of the proposed design by conducting several experiments on both multivariate synthetic and  real-world electroencephalogram (EEG) signals. The IMFs acquired  from the proposed hardware design have been validated in terms of their accuracy and execution time.

The proposed hardware for MEMD computation was implemented on Xilinx FPGA (Xilinx Virtex-7 FPGA VC707 Evaluation Kit) using Verilog programming. The Xilinx ISE Design Suite 14.7 was used to synthesize the Verilog code and the results were presented after post-synthesis, place and route procedure. In the proposed design, we used low-discrepancy Hammersley sequence to generate a set of $K=8$ directions vector for taking signal projections. The design was tested for $N=4$ number of input channels. The stopping criteria for the sifting process used $S=4$ number of iterations. Fixed-point data format of Q16.8 was used in our design.

\subsection{Case Study 1: Decomposition of Synthetic Data}
The proposed implementation for MEMD was evaluated on a synthetic quadri-variate (four channels) signal. Each channel was constructed from set of four pure tones: $f_1=50kHz$, $f_2=150kHz$, $f_3=350kHz$ and $f_4=800kHz$. One sinusoid signal ($f_3$) was kept common in all input channels. The $\bar{X}(t)$ is the resulting quadri-variate input signal consisting of $\{x_1(t),x_2(t),x_3(t),x_4(t)\}$ components, which are individually given as
\begin{equation} \label{eq9}
\begin{aligned}
&{x_{1}} =150 \sin ({{2\pi {f_{1}}t} })+150 \sin ({{2\pi {f_{3}}t}})+150 \sin ({{2\pi {f_{4}}t}}),\\
&x_2=150 \sin ({{2\pi {f_{1}}t} })+150 \sin ({{2\pi {f_{3}}t} }),\\
&x_3=150 \sin ({{2\pi {f_{2}}t} })+150 \sin ({{2\pi {f_{3}}t} })+150 \sin ({{2\pi {f_{4}}t} }),\\
&x_4=150 \sin ({{2\pi {f_{1}}t} })+150 \sin ({{2\pi {f_{2}}t} })+150 \sin ({{2\pi {f_{3}}t} }).
\end{aligned}
\end{equation}

\begin{figure*}[t]
	\centering
	\captionsetup{justification=centering}
	\hspace{-2mm}
	\includegraphics[height=1.5in,width=6.25in,trim=15mm 60mm 5mm 1mm, clip=true]{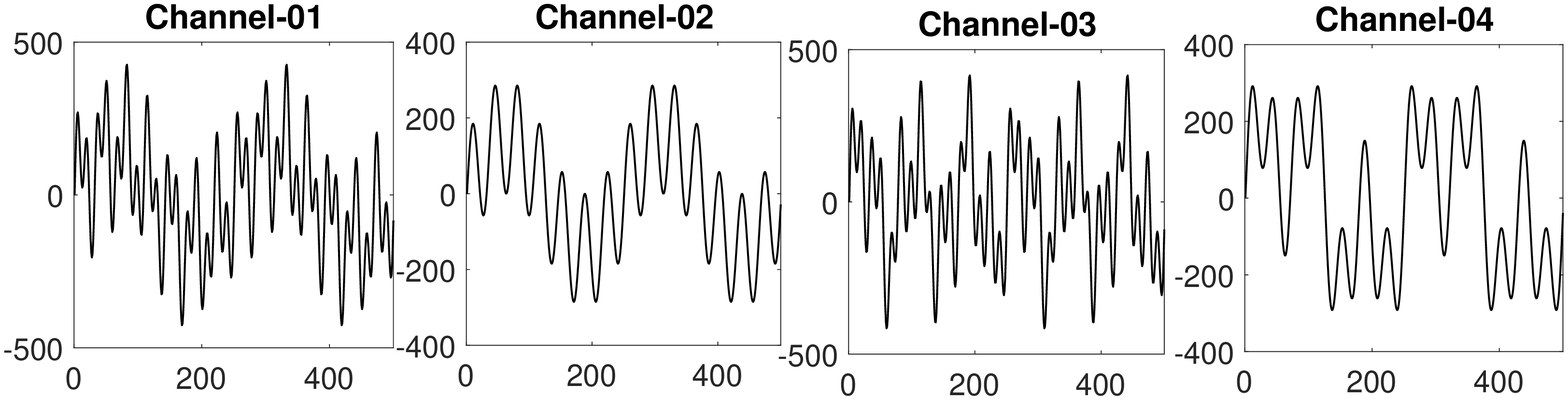}
	\caption{Synthetic quadri-variate input signal consisting of channels $x_1(t),x_2(t),x_3(t)$ and $x_4(t)$, from left to right.}
	\label{Fig_input_syn}       
\end{figure*}

The sampling frequency of the input signal was taken to be $F_s=30$ MHz. The input signal $\bar{X}(t)$ is shown in Fig. \ref{Fig_input_syn}.  

\begin{figure*}[t]
	\centering
	\captionsetup{justification=centering}
	\includegraphics[scale=0.575,trim=20mm 5mm 20mm 5mm, clip=true]{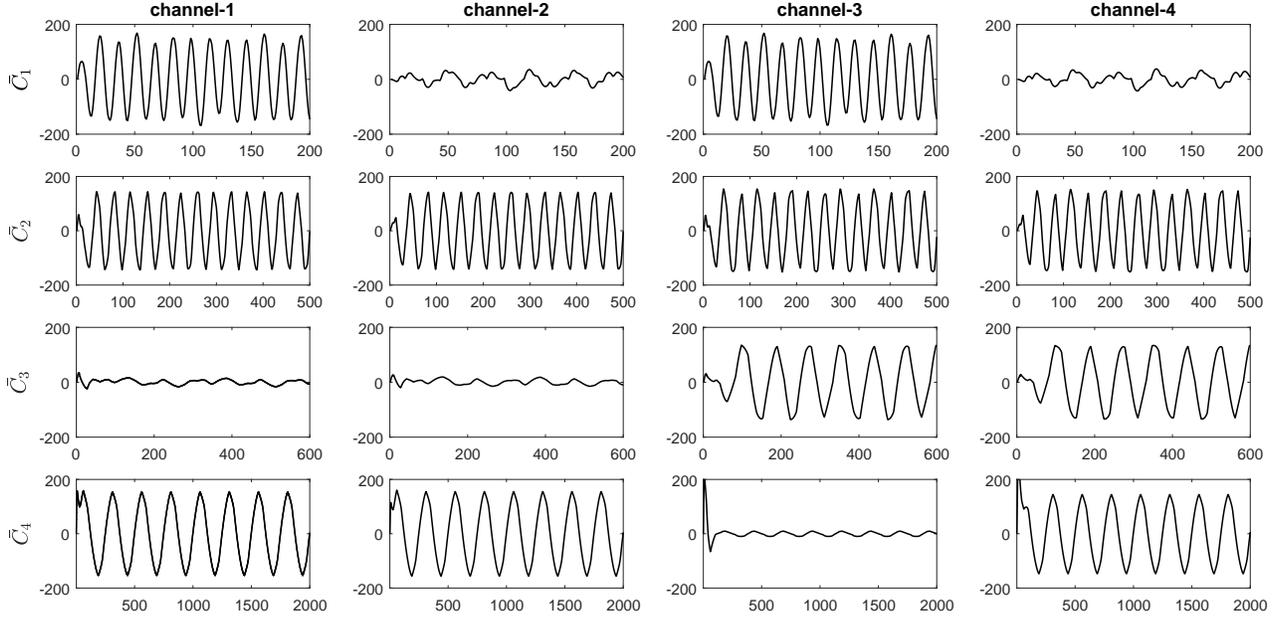}
	\caption{Decomposition of a the quadri-variate synthetic signal into multivariate IMFs $\bar{C_{1}}$, $\bar{C_{2}}$, $\bar{C_{3}}$ and $\bar{C_{4}}$ using the proposed MEMD based architecture.}
	\label{Fig_output_syn}       
\end{figure*}

The $\bar{X}(t)$ was given to the proposed MEMD design for the decomposition of the signal into $M=4$ IMFs. The resulting IMFs are shown in Fig. \ref{Fig_output_syn}. It can be noticed from the Fig. \ref{Fig_output_syn} that all original tones, with frequencies $f_1=50kHz$, $f_2=150kHz$, $f_3=350kHz$ and $f_4=800kHz$, have been decomposed accurately by the proposed architecture. We also see the signature mode-alignment property of MEMD \cite{x25}, \cite{x26} within the decomposition. Specifically, the $f_3=350kHz$ tone that was present in all channels has been decomposed as IMF2 within all channels. Moreover, the alignment of common frequency modes is evident in all remaining IMFs too.
%

\begin{table}[]
	\caption{Correlation of the decomposed IMF with ground truth (pure tones).}
	\label{tab_cor}
	\begin{tabular}{|c|c|c|c|c|}
		\hline
		\textbf{\begin{tabular}[c]{@{}c@{}}Correlation coefficient\\ (COR)\end{tabular}} &
		\textbf{\begin{tabular}[c]{@{}c@{}}Channel\\ (01)\end{tabular}}&
			\textbf{\begin{tabular}[c]{@{}c@{}}Channel\\ (02)\end{tabular}}&
			\textbf{\begin{tabular}[c]{@{}c@{}}Channel\\ (03)\end{tabular}}&
		     \textbf{\begin{tabular}[c]{@{}c@{}}Channel\\ (04)\end{tabular}} \\ \hline
		\begin{tabular}[c]{@{}c@{}}$\bar{C_{1}}(t)$\& \\ $f_4=800kHz$ \end{tabular} & 0.997 & 0.104 & 0.997 & 0.104 \\ \hline
		\begin{tabular}[c]{@{}c@{}}$\bar{C_{2}}(t)$ \&\\  $f_3=350kHz$\end{tabular} & 0.989 & 0.992 & 0.988 & 0.99 \\ \hline
		\begin{tabular}[c]{@{}c@{}}$\bar{C_{3}}(t)$ \&\\  $f_2=150kHz$ \end{tabular} & 0.005 & 0.112 & 0.936 & 0.936 \\ \hline
		\begin{tabular}[c]{@{}c@{}}$\bar{C_{4}}(t)$ \&\\  $f_1=50kHz$ \end{tabular} & 0.985 & 0.996 & 0.025 & 0.944 \\ \hline
	\end{tabular}
\end{table}

The accuracy of the decomposed IMFs had been verified by computing the correlation coefficient between each decomposed IMF and its corresponding ground truth (a pure tone). The correlation coefficient obtained between each pair (pure tone $\&$ the IMF of each channel) are presented in Table \ref{tab_cor}. It can be observed that those channels where we expected tones to appear have correlation values close to unity highlighting the accuracy of the decomposition. The remaining entries in the Table \ref{tab_cor} all have negligible correlation values as expected. The $f_3=350kHz$ tone that was present in all channels had correlation values greater than $0.9$ for all channels in $\bar{C_{2}}$.


\begin{figure}[h!]
	\centering
	\captionsetup{justification=centering}
	\includegraphics[height=1.42in,width=3.5in,trim=2mm 2mm 10mm 2mm,clip=true]{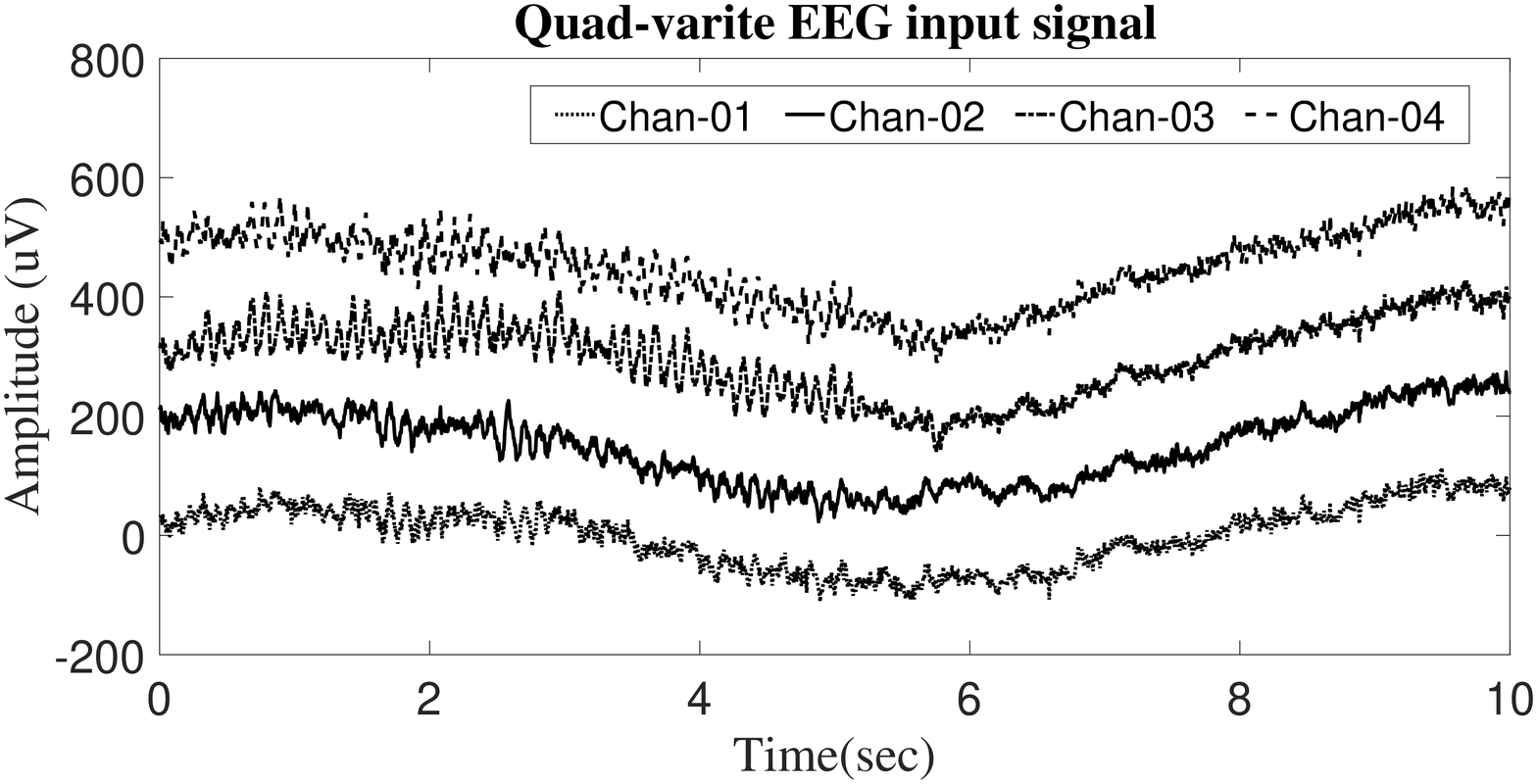}
	\caption{Time plot of four channel input EEG signal.}
	\label{EEG_in}       
\end{figure}

\subsection{ Case Study 2: Decomposition of multichannel electroencephalogram (EEG) data}
Owing to the nonstationary nature of real-world biomedical signals, data-driven multiscale approaches have found a lot of applications in biomedical signal processing. Moreover, a large class of biomedical signals, e.g., electrocardiogram (ECG) and electroencephalogram (EEG), are inherently multichannel. That makes MEMD a suitable choice to process such signals. 
To that end, our work is particularly relevant since it provides the first ever fully FPGA based on-line architecture for MEMD computation. Hence, those biomedical applications which demand on-line multiscale operation with high throughput and low latency requirements can benefit from our proposed architecture.  

Here, we demonstrate the application of our MEMD design architecture in multivariate EEG signal decomposition. The EEG data was collected to measure brain activity during two states: i) eyes-open and fully attentive state, ii) eyes-closed  and relaxed state. During the data collection, the subject remained in the relaxed states (eye-closed) for five seconds and was then asked to be in attentive state with eyes opened for the next five seconds.

The EEG data is known to exhibit alpha rhythm (dominant frequency component(s) in the range of 10$Hz$) during the relaxed state which is not present during the attentive (eyes open) state. A number of BCI related applications utilize those alpha rhythms to detect the relaxed state of the brain and multiscale data driven approaches have been widely used to achieve that.
 
\begin{figure*}[t]
	\centering
	\captionsetup{justification=centering}
	\includegraphics[scale=0.55,trim=1mm 1mm 12mm 0mm,clip=true]{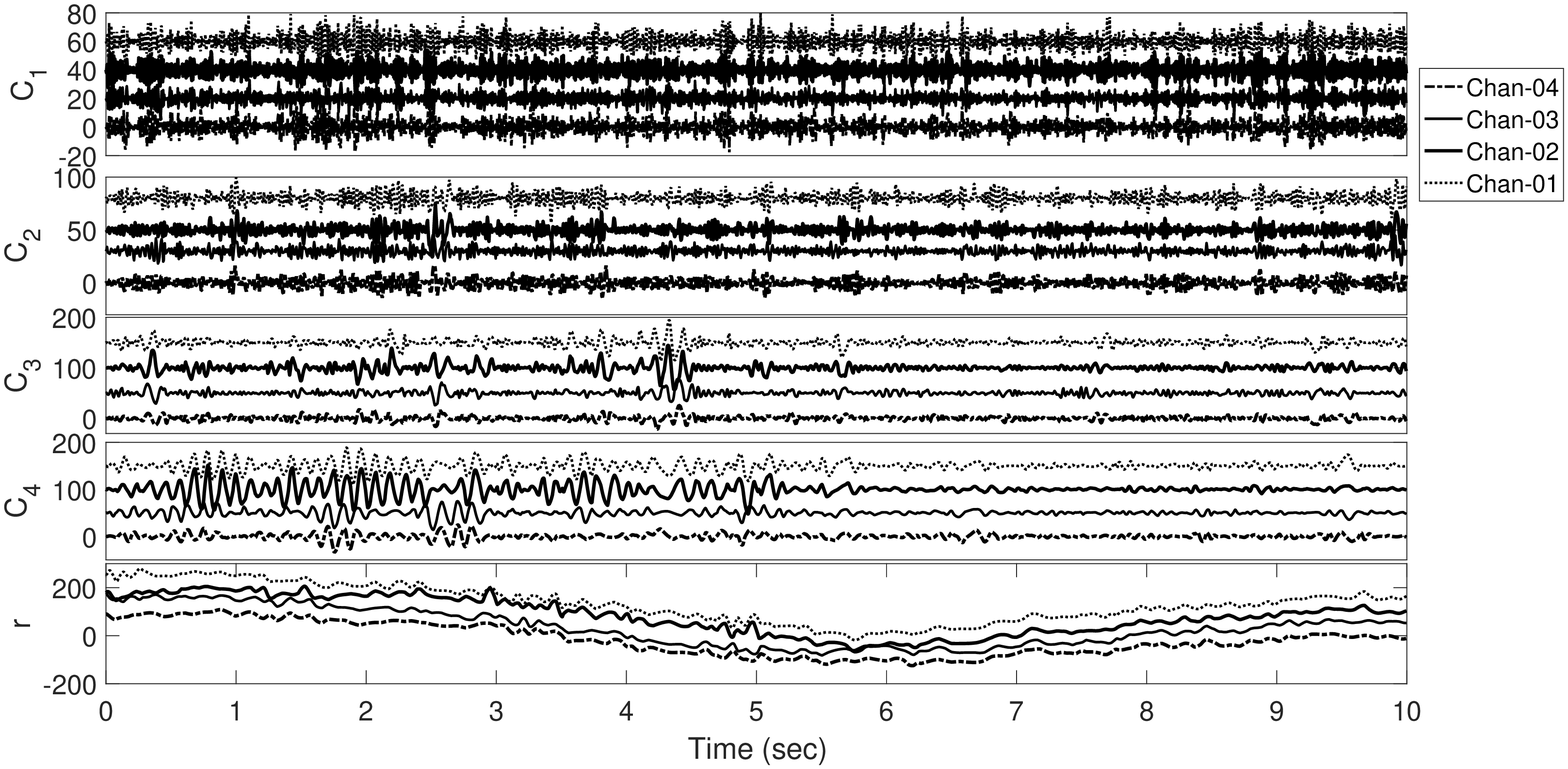}
	\caption{Decomposition of multivariate EEG signal into IMFs $C_1$, $C_2$, $C_3$, $C_4$ and residue $r$.}
	\label{EEG}
\end{figure*}

The Open BCI Cyton board was used to acquire the four-channel EEG signal that is shown in Fig. \ref{EEG_in}. The data was decomposed into $M=4$ IMFs and a residue for the purpose of analyzing eye-open and eye-closed activity. The aim here is to separate $\alpha$ rhythms (having frequency range of $8Hz$ to $15Hz$) from input EEG dataset by using the proposed architecture. The IMFs obtained from the MEMD along with the residue $r$ are shown in Fig. \ref{EEG}.

\begin{figure*}[!h]
	\centering
	\captionsetup{justification=centering}
	\hspace{-12mm}
	\includegraphics[scale=0.55,trim=2mm 2mm 2mm 2mm,clip=true]{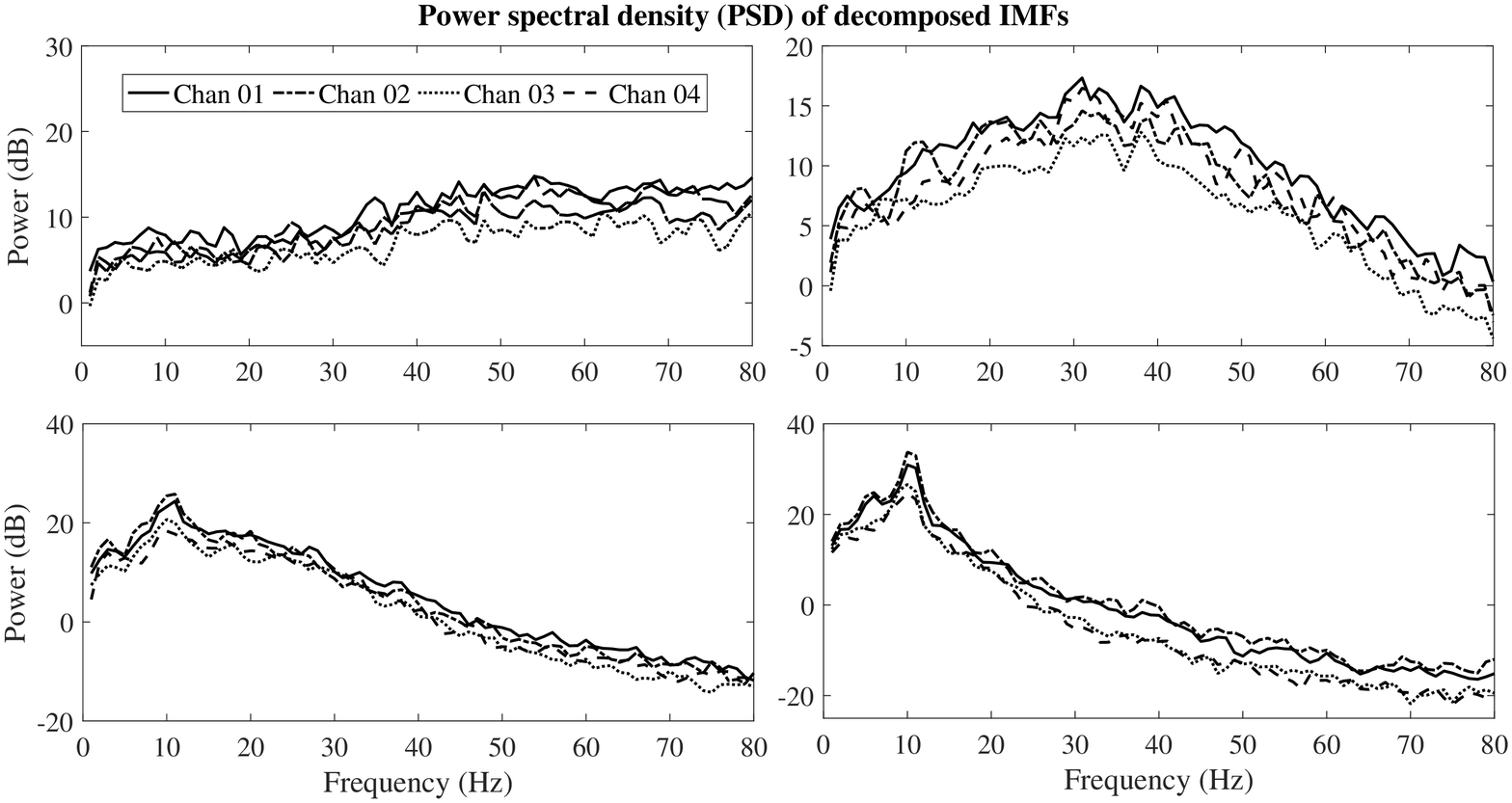}
	\caption{Power spectral density (PSD) of decomposed EEG signal ($C_1$ (top left), $C_2$ (top right), $C_3$ (lower left), and  $C_4$ (lower right)).}
	\label{EEG_IMF_freq}
\end{figure*}


It can be noticed from the figure that the $10Hz$ frequency component(s), corresponding to alpha rhythms, are decomposed in the fourth IMF ($C_4$) of each channel. The first IMF mostly picks up very high frequency components from EEG which may be present due to noise and/or other unexplained brain phenomenon. There are some cases of mode-mixing across IMFs, however, which may be present due to inherent limitations of the EMD-based approaches. The power spectral density (PSD) of the IMFs of each channel are also shown in Fig. \ref{EEG_IMF_freq} to validate our observations. It is evident from the PSD estimation that the fourth IMF $C_4$ shows a peak around $10Hz$ signal confirming its status as $\alpha$ wave. The overlapping of power spectra from multiple channels of same-index IMFs in Fig. \ref{EEG_IMF_freq} highlight the mode alignment property of MEMD. The ability of MEMD to i) separate nonstationary data into its constituent frequency components, and ii) aligning similar frequency components across multiple channels in same indexed IMFs, makes it a powerful and viable tool in several biomedical related applications.       
 
 \subsection{Hardware resources utilization and Timing analysis }
 
Hardware resources utilization and computational timing analysis of the proposed architecture is reported in this section. The prototype of the proposed architecture was developed using Xilinx Virtex-7 FPGA VC707 Evaluation platform (FPGA Virtex-7, XC7VX485T). The detail about hardware resources, the operating frequency and the slices used for the proposed architecture is reported in Table \ref{t2}. It can be noticed that the proposed design utilized 73\% slice LUTs, 11\% slice register and 10\% DSP 48EI slices from the available resourses.
\begin{table}[!h]
	\renewcommand{\arraystretch}{1.3}
	\caption{Hardware Utilization Of The Proposed Architecture Using Virtex-7, XC7VX485T-2FFG1761}
	\label{t2}
	\centering
	\begin{tabular}{|c|c|c|c|}
		\hline
		Logic Utilisation & Used & Available & Total Utilisation \\
		\hline 
		No. of Slice Registers & 66906	& 607200	&11$\%$ \\
		\hline 
		No. of Slice LUTs & 222538	& 303600	&73$\%$ \\
		\hline 
		No. of fully used LUT-FF pairs & 13930	&275514	&5$\%$ \\
		\hline 
		No. of bonded IOBs & 322	&700	&46$\%$ \\
		\hline 
		No. of BUFG/BUFGCTRLs & 3	&32	&9$\%$ \\
		\hline 
		No. of DSP 48E1s & 288	&2800	&10$\%$\\
		\hline 
	\end{tabular}
\end{table}

\begin{table}[h!]
	\renewcommand{\arraystretch}{1.3}
	\caption{Timing Analysis Of The Proposed Design}
	\label{t3}
	\centering
	\begin{tabular}{|c|c|} \hline 
		Sampling frequency/clock rate &  31 MHz \\ \hline 
		Initial delay for single iteration  & 0.65$\mu$s \\ \hline 
		Computational time for IMF1/1000 samples  & 36.5$\mu$s \\ \hline 
		Computational time for IMF2/1000 samples  & 40.70$\mu$s \\ \hline 
		Computational time for IMF3/1000 samples & 44.9$\mu$s \\ \hline 
		Computational time for IMF4/1000 samples & 49.1$\mu$s \\ \hline 
		Computational time for residue/1000 samples  & 49.1$\mu$s \\ \hline 
		Overall data throughput  & 1295 Mbps \\ \hline 
	\end{tabular}
\end{table}

The Table \ref{t3} gives timing analysis of the proposed architecture. Our design decomposes multivariate signals into $M=4$ IMFs with the maximum operating frequency of 31MHz. The design uses at least three extrema points to start the computation process, hence, there is an initial delay of 0.56$\mu s$ to fill the pipeline stages at the start. Moreover, each iteration of the sifting process is computed using a pipeline architecture that increases the throughput of the system. Similarly, IMFs are also computed in a pipelined fashion for every channel.

The computation time required to decomposition of first IMF for all channels was 36.5$\mu s$  for 1000 samples. Our proposed architecture only took 49.1$\mu s$ to decompose multivariate signals into four number of IMFs and a residue. {Moreover, the proposed design attained throughput of 1295 Mbps.}

\section{Discussion and Conclusions}\label{sec: conclusion}
We have proposed a parallel FPGA-based hardware architecture for multivariate extension of EMD algorithm. The design is suitable for on-line decomposition of multivariate signals in real-time. The best of our knowledge, no fully FPGA-based hardware for generic multivariate extension of EMD currently exists. For a specific case of bivariate signals, however, a parallel hardware design for computing bivariate extension of EMD (BEMD) was proposed in \cite {21}. Since our design can handle bivariate signals as a special case, it would be pertinent to compare our architecture with \cite {21}. The main algorithmic differences between BEMD and MEMD include: i) Unlike BEMD, MEMD can decompose more than two-channel signals; ii) projection vectors in MEMD are generated in N-dimensional spaces using low-discrepancy Hammersley sequence as opposed to uniform vectors in 2D in the case of BEMD.

\begin{table}[t]
	\caption{Comparison between state of art design}
	\label{tab:my-table}
		\begin{tabular}{|p{2.5cm}|p{2.5cm}|p{2.5cm}|}
			
		\hline
		& \begin{tabular}[c]{@{}l@{}}BEMD-Lin\\   \cite {21} \end{tabular} & \begin{tabular}[c]{@{}l@{}}Proposed\\   design\end{tabular} \\ \hline
		Cock rate (MHz) & 25 & 31 \\ \hline
		No of Input channel & 2 & 4 \\ \hline
		Direction vector & 8 & 8 \\ \hline
		Processing Time & 58.4$\mu$s for 1000 sample & 49.1$\mu$s for 1000 sample \\ \hline
		Throughput (Mbps) & 261.28 for STI & 1295 for CSI/STI \\ \hline
		Hardware Platform & \begin{tabular}[c]{@{}l@{}}FPGA\\   Xilinx Kintex-7,\\   (XC7K480T)\end{tabular} & \begin{tabular}[c]{@{}l@{}}FPGA\\   Xilinx Virtex-7\\   (XC7VX485T)\end{tabular} \\ \hline
		Interpolation technique & STI & STI, CSI \\ \hline
		Architecture & Pipeline & \begin{tabular}[c]{@{}l@{}}Pipeline\\   (STI, CSI)\end{tabular} \\ \hline
		Data format & Integer & Fixed point \\ \hline
	\end{tabular}
\end{table}

In terms of architectural design, the differences are as follows. Firstly, \cite{21} implemented linear interpolation scheme, also termed as sawtooth interpolation (STI), to generate envelopes within sifting process. Our proposed design for MEMD, on the other hand, implements a more accurate and well-established CSI scheme for the sifting process. Secondly, the design in \cite{21} used integer format that severely compromised its accuracy. Our architectural design of MEMD is based on fixed-point format which improves its overall precision and accuracy.   
The processing time of BEMD \cite{21} for 1000 input samples was 58.4$\mu$s whereas the proposed architecture required 49.1$\mu$s to decompose 1000 samples of a 4-channel data set. {This is because the proposed design uses look-up table for division operation to calculate spline and uses CSD based multiplication within the projection module.} The maximum achieved clock rate for the proposed architecture is 31MHz as compared to 25MHz for the BEMD architecture given in \cite{21}. 

As compared to the graphical processing unit (GPU) based implementation of MEMD \cite{x20}, the proposed FPGA-based hardware architecture offers the following advantages: i) delivery of high computational density per watt; ii) less power dissipation; iii) on-line and real time processing of multivariate data since the design in \cite{x20} can only perform batch processing. {GPU based system has one major disadvantage, which is challenging for online real-time applications. i.e., data transfer overhead from host system (CPU) to the memory of GPU. This data transfer overhead increases the overall execution time of the system.} All the above attributes make the FPGA based design a more viable option for on-line portable applications such as in biomedical engineering.

One limitation of the proposed architecture is that it could only use up to 8 direction vector within sifting process due to resource limitation on FPGA platforms. Parallel architectures with possibility of even higher number of direction vector can improve the accuracy of the design further; which can be possible to implement on ultra-scale version of the FPGAs.



\bibliographystyle{IEEEtr} 
\bibliography{main}

\begin{thebibliography}{10}

\bibitem{1}
C.~C. Tung, Z.~Shen, Q.~Zheng, S.~R. Long, N.-C. Yen, H.~H. Liu, N.~E. Huang,
  H.~H. Shih, and M.~C. Wu, ``{The empirical mode decomposition and the Hilbert
  spectrum for nonlinear and non-stationary time series analysis},'' {\em
  Proceedings of the Royal Society of London. Series A: Mathematical, Physical
  and Engineering Sciences}, vol.~454, no.~1971, pp.~903--995, 2002.

\bibitem{x2}
D.~Chen, D.~Li, M.~Xiong, H.~Bao, and X.~Li, ``{GPGPU-aided ensemble
  empirical-mode decomposition for EEG analysis during anesthesia},'' {\em IEEE
  Transactions on Information Technology in Biomedicine}, vol.~14, no.~6,
  pp.~1417--1427, 2010.

\bibitem{5}
A.~J. Nimunkar and W.~J. Tompkins, ``{EMD-based 60-Hz noise filtering of the
  ECG},'' in {\em Annual International Conference of the IEEE Engineering in
  Medicine and Biology - Proceedings}, pp.~1904--1907, 2007.

\bibitem{x4}
D.~S. Singh and Q.~Zhao, ``{Pseudo-fault signal assisted EMD for fault
  detection and isolation in rotating machines},'' {\em Mechanical Systems and
  Signal Processing}, vol.~81, pp.~202--218, 2016.

\bibitem{x5}
D.~Y.~Y. Li, C.~Rehtanz, and R.~Xiu, ``{Analysis of low frequency oscillations
  in power system based on HHT technique},'' in {\em 2010 9th Conference on
  Environment and Electrical Engineering, EEEIC 2010}, pp.~289--292, 2010.

\bibitem{x6}
N.~ur~Rehman, S.~Ehsan, S.~M.~U. Abdullah, M.~J. Akhtar, D.~P. Mandic, and
  K.~D. McDonald-Maier, ``{Multi-Scale pixel-based image fusion using
  multivariate empirical mode decomposition},'' {\em Sensors (Switzerland)},
  vol.~15, no.~5, pp.~10923--10947, 2015.

\bibitem{x7}
H.~Hao, H.~L. Wang, and N.~U. Rehman, ``{A joint framework for multivariate
  signal denoising using multivariate empirical mode decomposition},'' {\em
  Signal Processing}, vol.~135, pp.~263--273, 2017.

\bibitem{X8}
Z.~Yang and H.~Ren, ``Feature extraction and simulation of eeg signals during
  exercise-induced fatigue,'' {\em IEEE Access}, vol.~7, pp.~46389--46398,
  2019.

\bibitem{x9}
A.~Mert and A.~Akan, ``Emotion recognition from eeg signals by using
  multivariate empirical mode decomposition,'' {\em Pattern Analysis and
  Applications}, vol.~21, no.~1, pp.~81--89, 2018.

\bibitem{X10}
Y.~Lv, R.~Yuan, and G.~Song, ``Multivariate empirical mode decomposition and
  its application to fault diagnosis of rolling bearing,'' {\em Mechanical
  Systems and Signal Processing}, vol.~81, pp.~219--234, 2016.

\bibitem{x11}
A.~Y. Mutlu and S.~Aviyente, ``Multivariate empirical mode decomposition for
  quantifying multivariate phase synchronization,'' {\em EURASIP Journal on
  Advances in Signal Processing}, vol.~2011, pp.~1--13, 2011.

\bibitem{x16b}
D.~Looney and D.~P. Mandic, ``Multiscale image fusion using complex extensions
  of emd,'' {\em IEEE Transactions on Signal Processing}, vol.~57, no.~4,
  pp.~1626--1630, 2009.

\bibitem{x12}
N.~Rehman and D.~P. Mandic, ``Multivariate empirical mode decomposition,'' {\em
  Proceedings of the Royal Society A: Mathematical, Physical and Engineering
  Sciences}, vol.~466, no.~2117, pp.~1291--1302, 2010.

\bibitem{x13}
C.~Park, D.~Looney, N.~ur~Rehman, A.~Ahrabian, and D.~P. Mandic,
  ``Classification of motor imagery bci using multivariate empirical mode
  decomposition,'' {\em IEEE Transactions on neural systems and rehabilitation
  engineering}, vol.~21, no.~1, pp.~10--22, 2012.

\bibitem{x15}
H.~Azami, K.~Smith, and J.~Escudero, ``Memd-enhanced multivariate fuzzy entropy
  for the evaluation of complexity in biomedical signals,'' in {\em 2016 38th
  Annual International Conference of the IEEE Engineering in Medicine and
  Biology Society (EMBC)}, pp.~3761--3764, IEEE, 2016.

\bibitem{x14}
X.~Lang, D.~Zhong, L.~Xie, and J.~Chen, ``Application of improved multivariate
  empirical mode decomposition to plant-wide oscillations characterization,''
  in {\em 2017 6th International Symposium on Advanced Control of Industrial
  Processes (AdCONIP)}, pp.~601--606, IEEE, 2017.

\bibitem{den}
B.~K. N.~{Rehman} and K.~{Naveed}, ``Data-driven multivariate signal denoising
  using mahalanobis distance,'' {\em IEEE Signals Processing Letters}, vol.~29,
  no.~9, pp.~1408--1412, 2019.

\bibitem{17}
Y.-J. Chiu, P.-L. Lee, C.-M. Huang, M.-H. Lee, and K.-K. Shyu, ``{Hardware
  Implementation of EMD Using DSP and FPGA for Online Signal Processing},''
  {\em IEEE Transactions on Industrial Electronics}, vol.~58, no.~6,
  pp.~2473--2481, 2010.

\bibitem{18}
S.~Cagdas and A.~Celebi, ``{FPGA implementation of cubic spline interpolation
  method for empirical mode decomposition},'' in {\em 2012 20th Signal
  Processing and Communications Applications Conference (SIU)}, pp.~1--4, 2012.

\bibitem{19}
Y.~Y. Hong and Y.~Q. Bao, ``{FPGA implementation for real-time empirical mode
  decomposition},'' {\em IEEE Transactions on Instrumentation and Measurement},
  vol.~61, no.~12, pp.~3175--3184, 2012.

\bibitem{20}
N.~F. Chang, T.~C. Chen, C.~Y. Chiang, and L.~G. Chen, ``{On-line empirical
  mode decomposition biomedical microprocessor for Hilbert Huang transform},''
  in {\em 2011 IEEE Biomedical Circuits and Systems Conference, BioCAS 2011},
  pp.~420--423, 2011.

\bibitem{x24}
S.~{Gul}, M.~F. {Siddiqui}, and N.~{Ur Rehman}, ``{FPGA Based Real-Time
  Implementation of Online EMD With Fixed Point Architecture},'' {\em IEEE
  Access}, vol.~7, pp.~176565--176577, 2019.

\bibitem{21}
Q.~W. Malik, N.~ur~Rehman, S.~Gull, S.~Ehsan, and K.~D. McDonald-Maier,
  ``{FPGA-Based Real-Time Implementation of Bivariate Empirical Mode
  Decomposition},'' {\em Circuits, Systems, and Signal Processing}, vol.~38,
  pp.~118--137, jan 2019.

\bibitem{x20}
T.~Mujahid, A.~U. Rahman, and M.~M. Khan, ``{GPU-Accelerated Multivariate
  Empirical Mode Decomposition for Massive Neural Data Processing},'' {\em IEEE
  Access}, vol.~5, pp.~8691--8701, 2017.

\bibitem{x28}
G.~{Rilling}, P.~{Flandrin}, P.~{Goncalves}, and J.~M. {Lilly}, ``Bivariate
  empirical mode decomposition,'' {\em IEEE Signal Processing Letters},
  vol.~14, no.~12, pp.~936--939, 2007.

\bibitem{x25}
N.~U. Rehman, S.~Ehsan, S.~M.~U. Abdullah, M.~J. Akhtar, D.~P. Mandic, and
  K.~D. McDonald-Maier, ``Multi-scale pixel-based image fusion using
  multivariate empirical mode decomposition,'' {\em Sensors}, vol.~15, no.~5,
  pp.~10923--10947, 2015.

\bibitem{x26}
N.~ur~Rehman, Y.~Xia, and D.~P. Mandic, ``Application of multivariate empirical
  mode decomposition for seizure detection in eeg signals,'' in {\em 2010
  Annual International Conference of the IEEE Engineering in Medicine and
  Biology}, pp.~1650--1653, IEEE, 2010.

\bibitem{x26b}
X.~Zhao, T.~H. Patel, and M.~J. Zuo, ``Multivariate emd and full spectrum based
  condition monitoring for rotating machinery,'' {\em Mechanical Systems and
  Signal Processing}, vol.~27, pp.~712--728, 2012.

\bibitem{x22}
J.~Cui and W.~Freeden, ``Equidistribution on the sphere,'' {\em SIAM Journal on
  Scientific Computing}, vol.~18, no.~2, pp.~595--609, 1997.

\bibitem{x23}
N.~Rehman, M.~M. Khan, M.~I. Sohaib, M.~Jehanzaib, S.~Ehsan, and
  K.~McDonald-Maier, ``{Image fusion using multivariate and multidimensional
  EMD},'' in {\em 2014 IEEE International Conference on Image Processing, ICIP
  2014}, pp.~5112--5116, 2014.

\bibitem{24}
N.~E. Huang, M.~L.~C. Wu, S.~R. Long, S.~S. Shen, W.~Qu, P.~Gloersen, and K.~L.
  Fan, ``{A confidence limit for the empirical mode decomposition and Hilbert
  spectral analysis},'' {\em Proceedings of the Royal Society A: Mathematical,
  Physical and Engineering Sciences}, vol.~459, no.~2037, pp.~2317--2345, 2003.

\bibitem{27}
S.~Kotz, ``{Tridiagonal Matrix},'' {\em Encyclopedia of Statistical Sciences},
  2005.

\end{thebibliography}

\end{document}